\documentclass{ifacconf}

\usepackage{graphicx}      
\usepackage{natbib}        
\usepackage{graphicx}
\graphicspath{{Images/}} 
\usepackage{eso-pic} 
\usepackage{transparent}

\usepackage{amsmath}
\usepackage{bm}
\usepackage[overload]{empheq}  
\usepackage{amssymb}

\usepackage{tabularx}
\usepackage{longtable} 
\usepackage{colortbl}
\usepackage{multirow}
\usepackage{makecell} 
\usepackage{arydshln}

\usepackage{comment}

\usepackage{tikz}
\newcommand\copyrighttext{%
  \centering\footnotesize Accepted for publication at the IFAC World Congress 2023, July 9-14, 2023, Yokohama, Japan. \\
\copyright 2023 the authors. This work has been accepted to IFAC for publication under a Creative Commons Licence CC-BY-NC-ND. }
\newcommand\copyrightnotice{%
\begin{tikzpicture}[remember picture,overlay]
\node[anchor=south,yshift=-10pt] at (current page.north) {\fbox{\parbox{\dimexpr\textwidth-\fboxsep-\fboxrule\relax}{\copyrighttext}}};
\end{tikzpicture}%
}

\begin{document}
\begin{frontmatter}
\copyrightnotice
\vspace{-0.5cm}

\title{Non-Invasive Experimental Identification of a Single Particle Model for LiFePO\textsubscript{4} Cells} 


\author{Andrea Trivella,} 
\author{Matteo Corno,} 
\author{Stefano Radrizzani,}
\author{Sergio M. Savaresi}

\address{All the authors are with Dipartimento di Elettronica, Informazione e Bioingegneria, Politecnico di Milano, Milan, Italy\\ (e-mail: matteo.corno@ polimi.it).}

\begin{abstract}                
The rapid spread of Lithium-ions batteries (LiBs) for electric vehicles calls for the development of accurate physical models for Battery Management Systems (BMSs). In this work, the electrochemical Single Particle Model (SPM)  for a high-power LiFePO\textsubscript{4} cell is experimentally identified through a set of non-invasive tests (based on voltage-current measurements only). The SPM is identified through a two-step procedure in which the equilibrium potentials and the kinetics parameters are characterized sequentially. The proposed identification procedure is specifically tuned for LiFePO\textsubscript{4} chemistry, which is particularly challenging to model due to the non-linearity of its open circuit voltage (OCV) characteristic.
The identified SPM is compared with a second-order Equivalent Circuit Model (ECM) with State of Charge dependency. Models performance is compared on dynamic current profiles. They exhibit similar performance when discharge currents peak up to 1C (RMSE between simulation and measures smaller than 20 mV) while, increasing the discharge peaks up to 3C, ECM's performance significantly deteriorates while SPM maintains acceptable RMSE ($<$ 50 mV).

\end{abstract}


\end{frontmatter}


\section{Introduction}
\label{sec:introduction}
\vspace{-0.3cm}
Lithium-ion batteries (LiBs) are nowadays the most widespread technology to storage electric energy [\cite{n1}]. Their utilisation is well-established in a variety of applications, from portable electronics to medical and aerospace equipment, and they are considered a required technology for the ongoing transition to the Electric Vehicles (EVs). LiBs are superior to other electrochemical energy storage devices thanks to their higher power, capacity and flexibility, since by modifying the chemical composition of the electrodes, different characteristics can be achieved in terms of specific energy (\textit{i.e.}, the amount of energy that can be stored in a mass unit)  and specific power (\textit{i.e.}, how fast that energy can be delivered).\par 
Among the different technologies, LiFePO\textsubscript{4}-cathode cells are mainstream in the automotive field, due to their high power density and their superior safety. Today, they are employed in a wide range of traction applications such as passenger cars, buses, logistic vehicles and low speed electric vehicles. The main advantages of LiFePO\textsubscript{4} are their flat open-circuit voltage characteristics, the long cycle life, the abundance and low price of the raw materials, and the better environmental compatibility compared to other cathode materials, see \cite{lifepo}.
Considering their extensive use and the merits mentioned so far, LiFePO\textsubscript{4} are continuously gaining attention both in the academic research field and for industrial applications. 
For instance, some car manufacturers are shifting LiB technologies toward cheaper LiFePO\textsubscript{4} cells [\cite{tesla}]: Tesla intends to equip them on  electric vehicles with shorter ranges and lower prices, while Ford plans to use them for fleet-oriented trucks.
The necessity of safety standards, the need for high performance and the still-high cost of these devices, motivate the requirement for a Battery Management System (BMS). A modern BMS is a real-time monitoring device which carries on functions like the estimation of the State of Charge (SoC), \textit{i.e.}, the amount of charge remaining in the battery, and State of Health (SoH), \textit{i.e.}, the level of ageing of the LiB, overcharge and over-discharge avoidance, temperature control and so on. In order to operate efficiently, BMSs need to rely on mathematical models to replicate the cells' dynamics and estimate the state of the battery pack [\cite{lin2019modeling}], which is composed of hundreds to thousands of cells. 
\par The main modeling techniques for Li-ion cells are Equivalent Circuit Models (ECMs) and Electrochemical Models (EMs). The first ones mimic the behaviour of the cell with a fictitious equivalent circuit, they are simple to implement and easy to calibrate. The EMs, on the other hand, describe the cell's dynamics through first principles and physical laws, they are superior in accuracy, but they are also more complex, and require more efforts to be calibrated and computed. The great advantage of EMs is their capability to provide a model also for the cell's internal states, such as the lithium concentration or the potential of the single electrodes. The Pseudo Two Dimensional (P2D), developed by \cite{DFN}, is the most complete EM model. P2D describes the cell's state evolution by modeling 1) the transport of the lithium ions in the electrolyte and 2) the diffusion of the lithium inside the electrode particles. The identification of P2D is challenging, due to its large number of parameters, its simulation is computationally demanding and calls for efficient discretization techniques [\cite{corno2014electrochemical}, \cite{corno2020efficient}]. The Single Particle Model (SPM), introduced in \cite{zhang1}, is a reduction of the P2D and offers a good compromise between accuracy and complexity: it neglects the electrolyte ions migration and relies on the solid phase diffusion dynamics.
On-board BMSs typically make use of simple ECMs to estimate the battery pack's SoC and SoH [\cite{lin2019modeling}], however, these models tend to lose their validity when high discharge rates are applied. The use of EMs in EVs applications can bring undeniable advantages: thanks to their physical insight, they offer a superior accuracy both in the estimation of the internal states and in the prediction of possible faults, optimizing the use of the cells and extending the useful life of the entire pack, say \cite{bms3}.
\par
The parameters identification for electrochemical models is a relatively recent subject of research. As pointed out in the review by \cite{laue}, electrochemical models are typically parameterized through a two-step procedure: equilibrium potential curves and kinetics parameters are found sequentially. The number of identified kinetic parameters and the entity of the experiments is variable. \cite{jon} and \cite{masoudi}, for example, identify a subset of most-sensitive kinetics parameters with pulse or C-Rate experiments, and maintain nominal values for the remaining ones. \cite{biz} perform a re-parametrization of the SPM and identify a reduced set of grouped parameters via Electrochemical Impedance Spectroscopy (EIS). \cite{plett}, instead, identify all the kinetics parameters of a P2D through a complex 4-step protocol, with synthetic data produced by \textsc{Comsol} P2D simulator. Equilibrium curves, on the other hand, are typically analytical functions taken from literature and directly parametrized using OCV measurments, see, for example \cite{reddy}, \cite{namor} or \cite{stefanopo}.
Other works rely on highly invasive experiments, like destructive half-cell measurements, \textit{e.g.}, \cite{halfcell}.
The objective of this study is the development of an identification procedure, through non-invasive IV (current-voltage) measurements, of an electrochemical model for LiFePO\textsubscript{4} high-power commercial cell.

The first key contribution of this work is development of a recursive procedure to estimate the cell Equilibrium Potentials with high accuracy: starting from literature analytical formulations, the curves are significantly improved with sequential local electrode-specific corrections  until error rates between model and OCV measures smaller than 10mV are achieved. After the identification of the remaining kinetics parameters, entrusted with a simulation error-based optimization routine, an extensive validation campaign is carried out. Thus, the second key contribution is the comparison, made with many and varied experiments at different C-Rates, of the identified SPM with an ECM, which is a benchmark for the modeling of LiBs. The electrochemical model appears to be superior then ECMs, especially increasing the discharge rates.

  \par The remainder of this paper is organized as follows: in Section \ref{s2}, the SPM is presented and identified, Section \ref{s3} is a quick overview of the identification protocol for the ECM, and in Section \ref{s4}, the results of the experimental validation and comparison are shown.

\section{SINGLE PARTICLE MODEL\\ IDENTIFICATION}  
\label{s2}
In this section, the Single Particle Model is presented and the proposed identification procedure is discussed.
\vspace{-0.2cm}
\subsection{Single Particle Model Overview}
\begin{figure}[!ht]
\hspace{0cm}
 \includegraphics[scale=0.28]{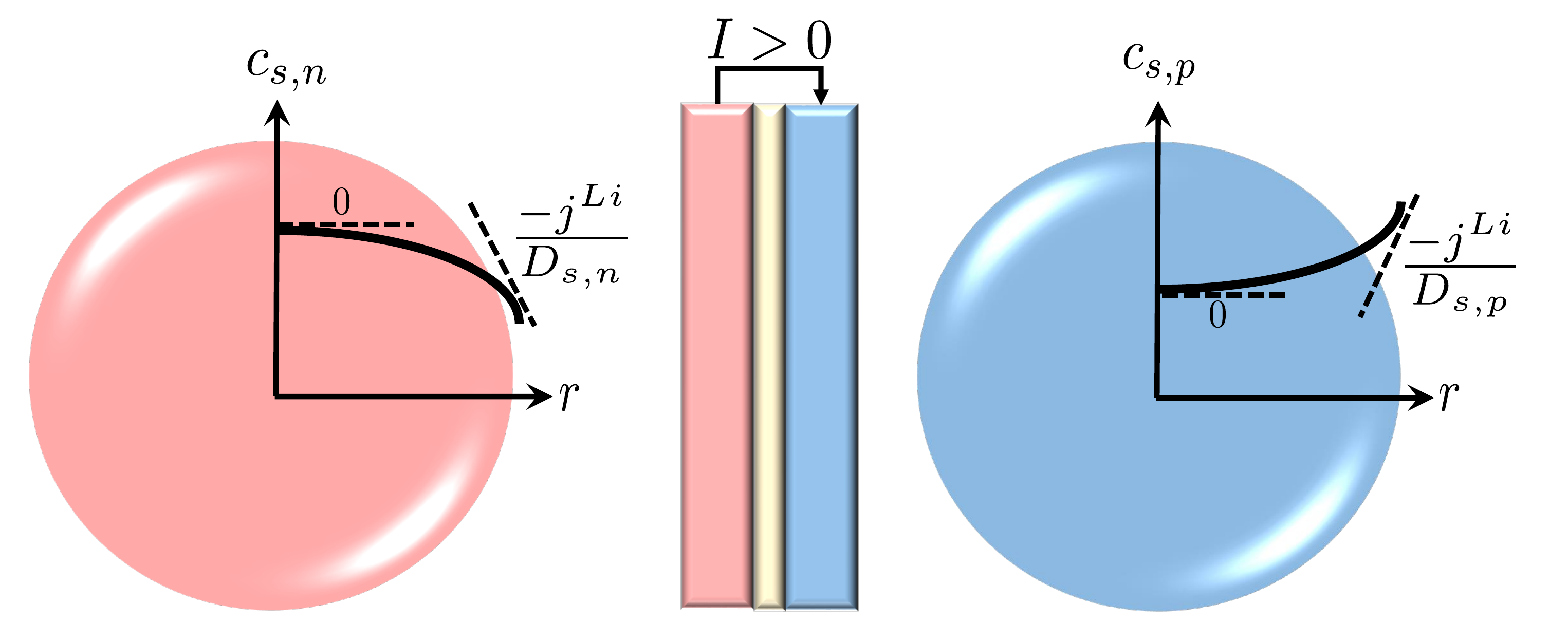}
 \vspace{-0.2cm}
    \caption{Illustrative example of lithium concentration distributions in the spherical particles during a discharge.}
    \label{fick}
    \vspace{-0.2cm}
\end{figure}

\vspace{-0.3cm}
 In the SPM, each electrode is modeled as a single particle with a variable lithium concentration, defined by the \textit{Fick's Law} and its boundary conditions [\cite{DFN}]:
\begin{equation}
    \frac{\partial c_{s,i}}{\partial t} = \frac{D_{s,i}}{r^2} \frac{\partial}{\partial r} \left(r^2 \frac{\partial c_{s,i}}{\partial r}\right),
    \label{fickk}
\end{equation}
\begin{equation}
   \text{ with }  \frac{\partial c_{s,i}}{\partial r}\Big|_{r=0} = 0 \text{ and } D_{s,i}\frac{\partial c_{s,i}}{\partial r}\Big|_{r=r_{s,i}} = -j^{Li},
\end{equation}
where $r$ is the spherical coordinate, $c_{s,i}$ is the solid-phase (s-p) concentration in the $i=\{n,p\}$ electrode, $D_{s,i}$ is the s-p diffusivity, $r_{s,i}$ is the particle's radius and $j_i^{Li}$ is the molar flux at the particle's surface.
\noindent The molar flux $j_i^{Li}$ is a function of input current, defined as:
\begin{equation}
j_n^{Li}=\frac{I\,r_{s,n}}{3\,V_n\,F} \text{ and } j_p^{Li}=-\frac{I\,r_{s,p}}{3\,V_p\,F},
    \label{flux2}
\end{equation}
where $I$ is the input current, $F$ is the Faraday's constant and $V_i$ is the electrode's volume, defined as:
\begin{equation}
    V_i = \varepsilon_i A_i \delta_i,
\end{equation}
where $\varepsilon_i$ is the electrode's volume fraction, $A_i$ is the electrode's area and $\delta_i$ is the electrode's thickness.\\
Fig. \ref{fick} shows a possible concentration distribution in anode (left) and cathode (right) during a discharge.

The cell terminal voltage is the difference between the equilibrium potentials $U_i(c_{s,i})$, major voltage contributions, intrinsic characteristics of each electrode chemistry, and overpotentials $\eta_i$, minor contributions arising when a current is applied to the cell's terminals, and a resistive drop:
\begin{equation}
   V = \left(U_p - \eta_p \right) -\left(U_n - \eta_n \right) - R_fI,
\end{equation}
where $R_f$ is the film resistance.
The equilibrium potentials are nonlinear functions of the surface stoichiometries $U_i=f^{NL}(\theta_{ss,i})$ where
 \begin{equation}
 \theta_{ss,i}=\frac{c_{s,i}}{c_{s,i}^{\mathrm{max}}}\Big|_{r=r_{s,i}}
 \end{equation}
is defined between the stoichiometry limits $\theta_{s,i}^{100\%}$ and $\theta_{s,i}^{0\%}$, which represent the percentage of the maximum lithium concentration reached when the cell is fully charged or discharged.
Finally, the overpotentials $\eta_i$ are defined by the \textit{Butler-Volmer} equation:
\begin{equation}
    \eta_i = \frac{2R_gT}{F}\sinh^{-1}{\left(\frac{F}{2i_{0,i}}j_i^{Li}\right)},
    \label{n1}
\end{equation}
where $R_g$ is the gas constant, $T$ is the temperature and
\begin{equation}
i_0 = k_i\sqrt{c_e^{\mathrm{avg}}\left(c_{s,i}^{\mathrm{max}}-c_{ss,i}\right)c_{ss,i}}.
\label{i0}
\end{equation}
 In \eqref{i0}, $k_i$ is the reaction rate constant and $c^\mathrm{avg}_e$ and $c_{s,i}^\mathrm{max}$ are the average electrolyte concentration and the maximum s-p concentration.
 In this work, the SoC is calculated as:
 \begin{equation}
    \mathrm{SoC} = \mathrm{SoC}_0 - \int_0^t\dfrac{I(t)}{3600\,Q_{\mathrm{nom}}}\mathrm{d}t
    \label{soc}
\end{equation}
where $Q_{\mathrm{nom}}$ is the cell's nominal capacity (3.2 Ah).
\par To numerically compute the \textit{Fick's Law}, it is discretized along the radial coordinate. For this purpose, $N=20$ points are defined on the half-spheres (the concentration in the other half is specular), reducing the original PDE to a set of ODEs in the time-domain.
 The approximation of the PDE is achieved via \textit{Chebyshev Orthogonal Collocation} [\cite{matlab_diff}], a spectral approach consisting in approximating the solution of the PDE by a sum of polynomials (\textit{Chebyshev polynomials}) that interpolate the solid concentration curves at the $N$ nodes (\textit{Chebyshev nodes}). 
 \vspace{-0.3cm}
\subsection{Single Particle Model Identification}
\begin{figure}[h!]
\centering
 \vspace{-0cm}
 \includegraphics[scale=1]{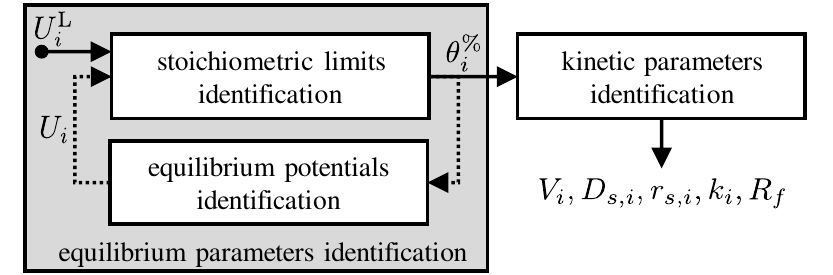}
 \vspace{-0.2cm}
    \caption{Identification procedure outline.}
    \label{steps}
    \vspace{-0.2cm}
\end{figure}

\vspace{-0.3cm}
The SPM depends on many parameters, whose identification is nontrivial.
The proposed identification procedure consists of two phases as shown in Fig. \ref{steps}.
 In the first step, the equilibrium potentials and the stoichiometry limits are identified exploiting OCV measurements. These curves model the cell potentials in static open circuit conditions, and can be retrieved regardless of the kinetics parameters ($V_i,\,r_{s,i},\,D_{s,i}$ ...). The open circuit model obtained in the first stage is then exploited to identify the kinetics parameters. These latter, characterizing the SPM dynamics, are identified applying dynamic current profiles.
\par \textit{Equilibrium Parameters}.
Given that graphite (LiC\textsubscript{6}) and LiFePO\textsubscript{4} are well-known materials, several analytical formulations of $U_n$ and $U_p$ can be found in the literature. However, the literature curves lead to inaccuracies and mismatches between predicted and measured OCV points. For this reason, the strategy adopted starts from a set of tentative curves ($U_i^\mathrm{L}$), then use them to find the stoichiometry limits, and finally refine them ($U^*_i$) to fit the OCV measures (see Fig. \ref{steps}).
\par
\begin{figure}[h!]
\vspace{-0.2cm}
\hspace{0.7cm}
 \includegraphics[scale=0.525]{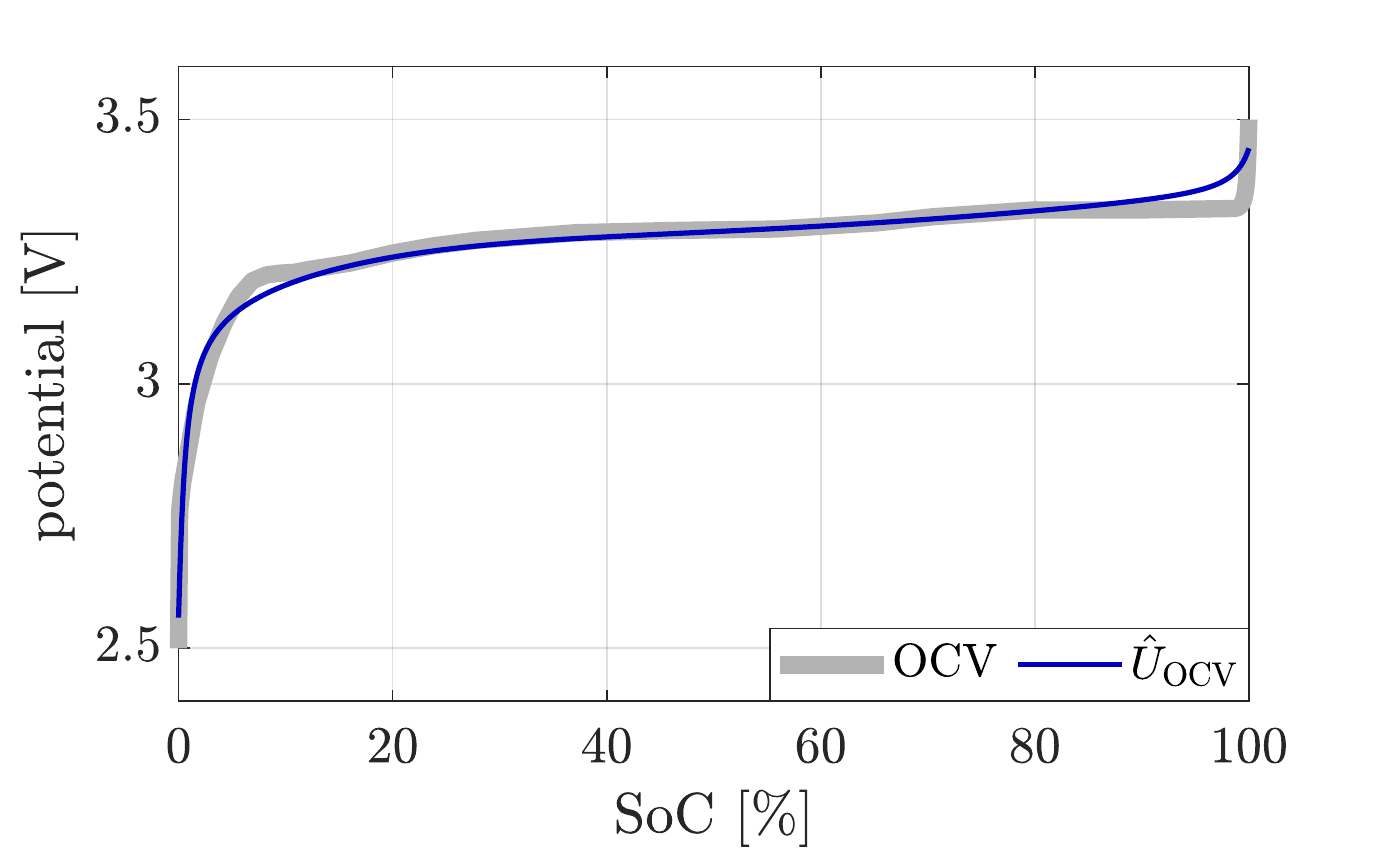}
 \vspace{-0.2cm}
    \caption{Equilibrium potential of SPM, obtained as difference of literature curves.}
    \label{ocvv}
    \vspace{-0cm}
\end{figure}

  The equilibrium curves have been defined as:
\begin{equation}
U_n^{*}=U^\mathrm{L}_n + E_n \text{  and  } U_p^{*}=U^\mathrm{L}_p + E_p.
\end{equation}
where $U_i^\mathrm{L}$ are taken form the literature, respectively from \cite{jon} and \cite{albi}:
\begin{equation}
    \centering
    U_{n}^\mathrm{L}(\theta_n) = \sum\limits_{i=0}^5 \gamma_{i}\theta_n^{\frac{i-2}{2}} +\sum\limits_{i=6}^7\gamma_{i}\exp\left( \gamma_{i+1}\theta_n\right),
\end{equation}
\begin{equation}
    \centering
    U_{p}^\mathrm{L}(\theta_p) = \gamma_{p,0} + \sum\limits_{i=1,4} \gamma_{p,i}\text{atan}\left(\gamma_{p,i+1}+\gamma_{p,i+2}\theta_p\right).
    \label{Uplit}
\end{equation}
$E_n$ and $E_p$ are, instead, a sum-of-gaussians and a sum-of-exponentials and are designed in a second step to locally improve the fitting of our model with to OCV values.\\ As a first step, we identify the stoichiometry limits on the curves $U^\mathrm{L}_n$ and $U^\mathrm{L}_p$, using OCV measurements from a Pulse Discharge Test (PDT), consisting of a complete cell discharge with current pulses, alternated by resting periods. In fact, when no currents are applied on the SPM model and transients are over (\textit{i.e.} in open circuit condition), we have:
\begin{equation}
      V \xrightarrow[t \to \infty]{} U_p - U_n = \mathrm{OCV}.
\end{equation}
In addition, a degree of freedom is introduced, allowing small curve translations in the Voltage-Stoichiometry domain, in order to improve the model convergence to the OCV  measures, rewriting $U_i = U_i(\theta_{ss,i}+a_i) + b_i$.\\
The optimal vector:
\begin{equation}
        x = \left[ \theta_{n}^{0\%},\, \theta_{p}^{0\%},\,\theta_{n}^{100\%},\,\theta_{p}^{100\%}, \, a_n,\,b_n,\,a_p,\,b_p \right]
        \label{problem}
\end{equation}
is found minimizing the error, in $j=1...$M points in the SoC domain, between the measures $\mathrm{OCV}(j)$ and the predictions $\Hat{U}_{\mathrm{OCV}}(x,j) = U_{p}(x,\theta_p(j))-U_{n}(x,\theta_n(j))$. \\
The optimization problem is therefore:
\vspace{-0.2cm}
\begin{equation}
\begin{aligned}
& \underset{\text{\large{\textit{x}}}}{\text{{minimize}}}
& &  \sum\limits_{j=1}^M \left[\mathrm{OCV}(j)-\Hat{U}_{\mathrm{OCV}}(x,j)\right]^2 \\
& \text{subject to}
& &  0 \leq  \theta_n^{0\%} \leq  \theta_n^{100\%} \leq 1 \\
& & & 0 \leq  \theta_p^{100\%} \leq  \theta_p^{0\%} \leq 1
\end{aligned}
\end{equation}
where the $\theta_i(j)$ are computed (\textit{e.g.} in the anode) exploiting the following relationship, necessary to move from the SoC domain to the stoichiometry domain and viceversa:
\begin{equation}
\theta_n(j) = \theta_n^{0\%} + \mathrm{SoC}(j)\left(\theta_n^{100\%}-\theta_n^{0\%}\right).
\label{socSPM}
\end{equation}
This nonlinear constrained problem is solved via \textsc{Matlab} function \verb+fmincon+ with interior-point method. The optimal limits and translations are:
\begin{align*}
    {x}^* = [0.007,\; 0.9821,\; 0.8185,\; 0.06,\\\;0.058,\;0.018,\;0.011,\;-0.018 ].
\end{align*}
We define the obtained set of stoichiometry limits as $\theta_0$.\\
The resulting open circuit potential, shown in Fig. \ref{ocvv}, has a RMSE of 19.3 mV.

\begin{figure}[h!]
\vspace{-0.2cm}
\hspace{0.6cm}
\vspace{-0.2cm}
 \includegraphics[scale=0.55]{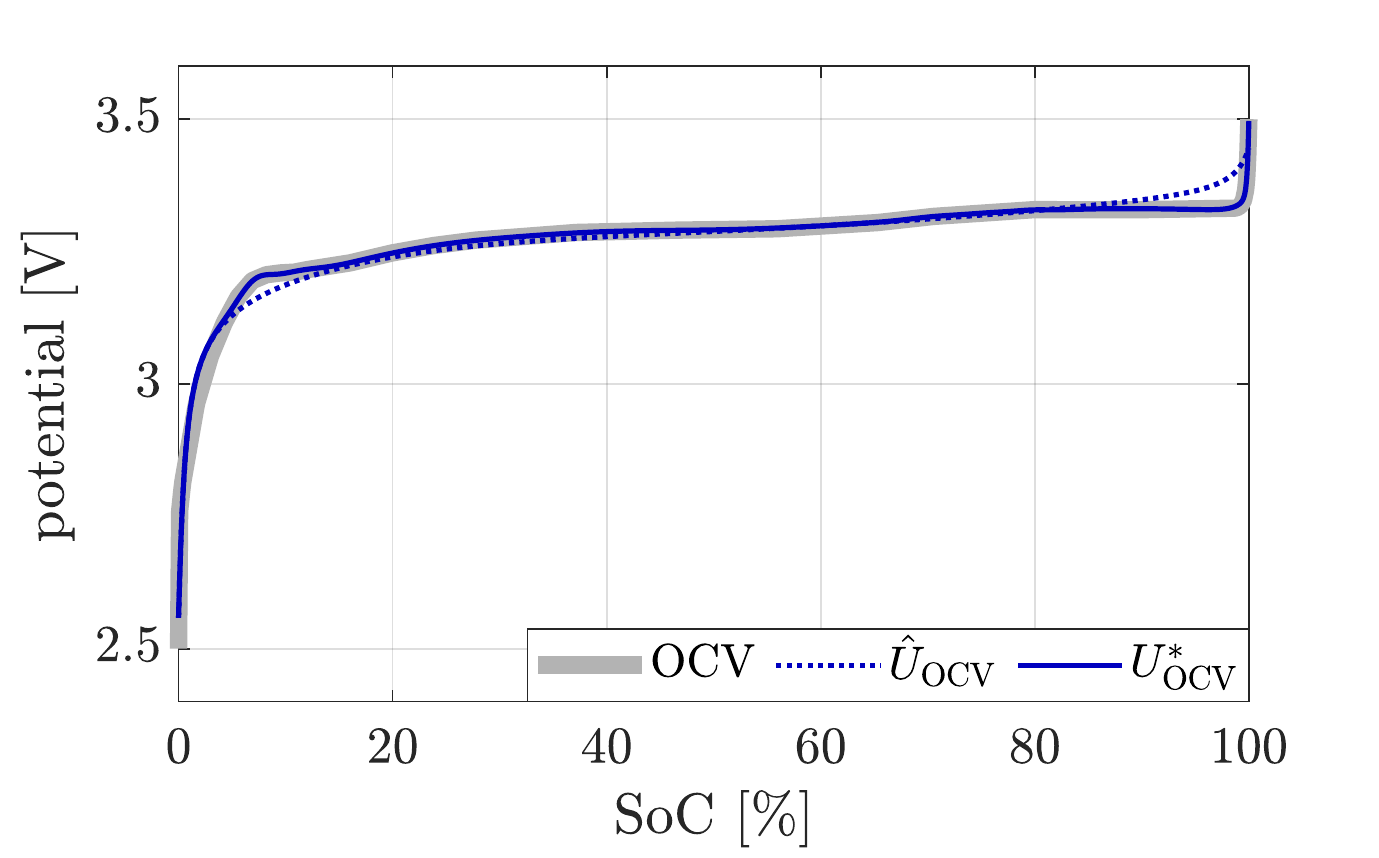}
    \caption{Final equilibrium potential of SPM, obtained as difference of literature curves locally refined.}
    \label{ocvfinal}
\end{figure}

From Fig. \ref{ocvv}, the identified model at this stage shows discrepancies for low and high SoC. Acting on the global shape of $U_i$ to correct local inaccuracies has proven to be inefficient and leading to the rising of new sources of error; thus, a sequential and local approach is preferred: we select an error zone, attribute that error to one electrode potential, and fit it locally using proper functions.
The ``bumps" in the low-SoC zone, for example, are typical of the anode's potential curve, and thus the relative residuals are modeled with a sum-of-gaussians, parametric in $\alpha$: 
\begin{equation}
\centering
    E_n = \sum\limits_{i=1}^6\alpha_{1,i}\exp\left(-\left(\dfrac{\alpha_{2,i}-\mathrm{SoC}}{\alpha_{3,i}}\right)^2\right)
    \label{equah}
\end{equation}
and added to the anode potential $U^\mathrm{L}_n$.
\noindent The high-SoC error is fitted through a sum-of-exponentials, parametric in $\beta$:
\begin{equation}
 \centering
E_p = \beta_1\exp\left( \beta_2\mathrm{SoC}\right)+\beta_3\exp\left( \beta_4\mathrm{SoC}\right)
\label{err1exp}
\end{equation}
and used to correct the cathode potential curve.
\noindent The final potential is computed as $U_{\mathrm{OCV}}^*=U^*_p - U^*_n$.
The resulting equilibrium potential difference is compared with the measured OCV in Fig. \ref{ocvfinal}, and exhibits an RMSE of 9.5mV (was 19.3 mV before the refinements).

\begin{figure}[h!]
 \vspace{-0.2cm}
 \hspace{-0.4cm}
 \includegraphics[scale=0.58]{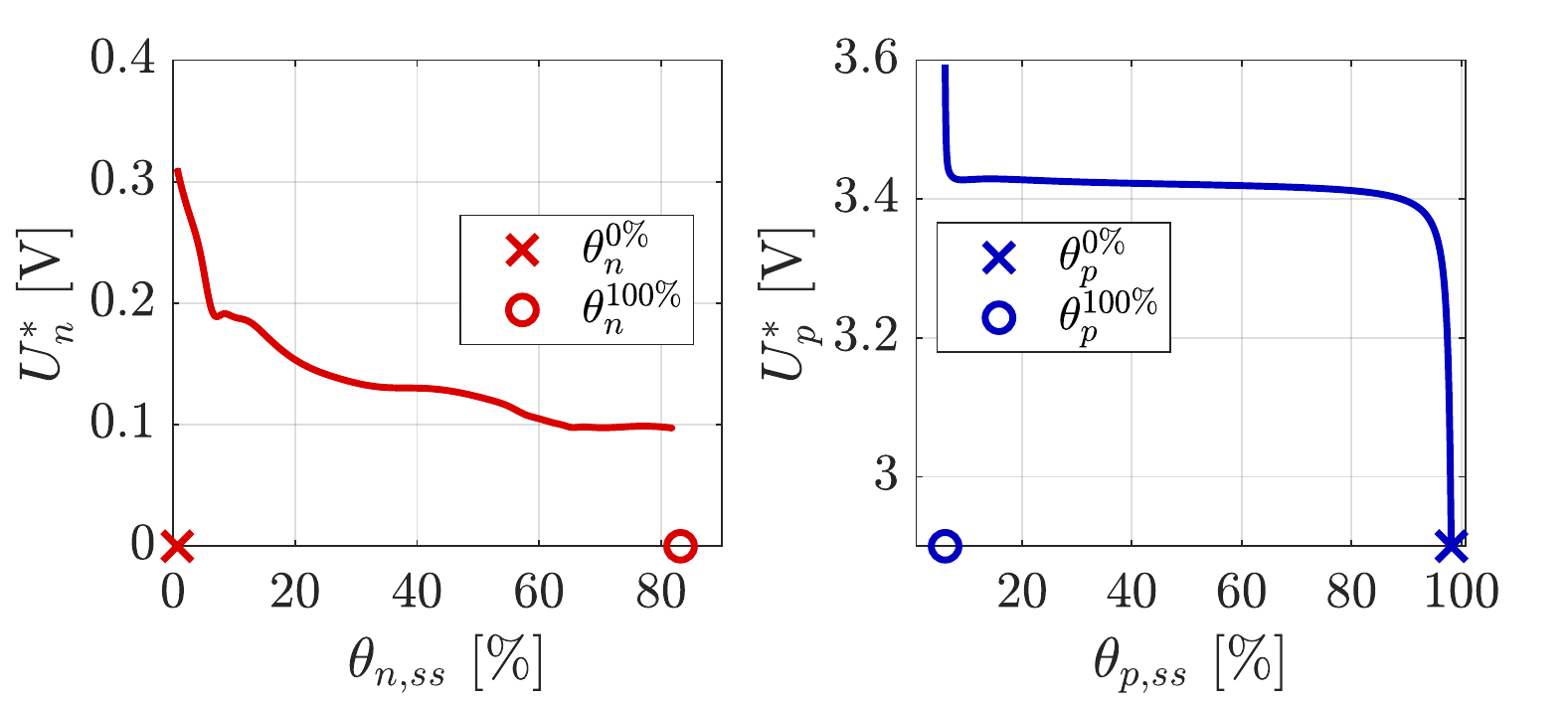}
 \vspace{-0.4cm}
    \caption{Equilibrium potentials of anode (red) and cathode (green) and relative stoichiometry limits.}
    \label{upun}
    \vspace{-0cm}
\end{figure}

As a final step, the stoichiometry limits must be re-optimized on the new set of refined curves. Problem \eqref{problem} is adjusted substituting $\hat{U}_{\mathrm{OCV}}$ with $U^*_{\mathrm{OCV}}$, and solved to find the new final limits $\theta^*$:

\begin{table}[h!]
\renewcommand{\arraystretch}{1.5}
\centering
\normalsize
\caption{Optimal stoichiometry limits.}
\begin{tabular}{ccccc}
\hline
parameter set & $\theta_{n}^{0\%}$ & $\theta_{p}^{0\%}$ & $\theta_{n}^{100\%}$ & $\theta_{p}^{100\%}$\\ 
 \hline
   $\theta^*$           &  0.0072     &    0.9822  & 0.8320 & 0.06\\  
   $\theta_0$        &  0.007     &    0.9821  & 0.8185 & 0.06    \\ 
\hline
\end{tabular}
\label{r0_par}
\end{table}
\noindent Anode's and Cathode's optimal equilibrium potentials are shown in Fig. \ref{upun}.

\par
\textit{Kinetic Parameters}.
The second phase consists of an identification of the kinetics parameters:
\vspace{0cm}
\begin{align*}
    x = \left[V_n,\,V_p,\,r_n,\,r_p,\,D_{s,n},\,D_{s,p},\,k_n,\,k_p,\,R_f \right].
\end{align*}  
Parameters $c_{s,i}^{\mathrm{max}}$ and $c_e^{\mathrm{avg}}$ have found to be consistent among different sources (e.g. \cite{par1}, \cite{par2} and \cite{zhang}) , and thus they are excluded from the identification.
The solver is initialized with the set of parameters referred to a LiFePO\textsubscript{4} chemistry [\cite{zhang}], and some bounds $[L_b,U_b]$ are used to constrain the solution to a realistic subset of values.\\
The identification dataset is composed of two discharge tests: a constant 1C discharge and a white noise (WN) discharge.\\ The kinetics parameters are optimized solving the least-square problem:
\begin{align*}
\begin{aligned}
& \underset{\text{\large{\textit{x}}}}{\text{{minimize}}}
& &  F_1(x)^TF_1(x)+F_2(x)^TF_2(x) \\
& \text{subject to}
& & L_b\leq x\leq U_b \\
& & & \theta_n^{0\%} \leq \theta_n(x) \leq  \theta_n^{100\%} \\
& & & \theta_p^{100\%} \leq \theta_p(x) \leq \theta_p^{0\%}
\end{aligned}
\end{align*}
where $F_i(x)=\tilde{V}-\hat{V}(x)$ is the difference between the measured and the simulated voltage of test $i={1,2}$.\\
\noindent The optimal values, obtained via \verb+fmincon+, are listed in Table \ref{optKIN}.

\begin{figure}[h!]
\vspace{-0.4cm}
\hspace{0.6cm}
\vspace{-0.2cm}
 \includegraphics[scale=0.55]{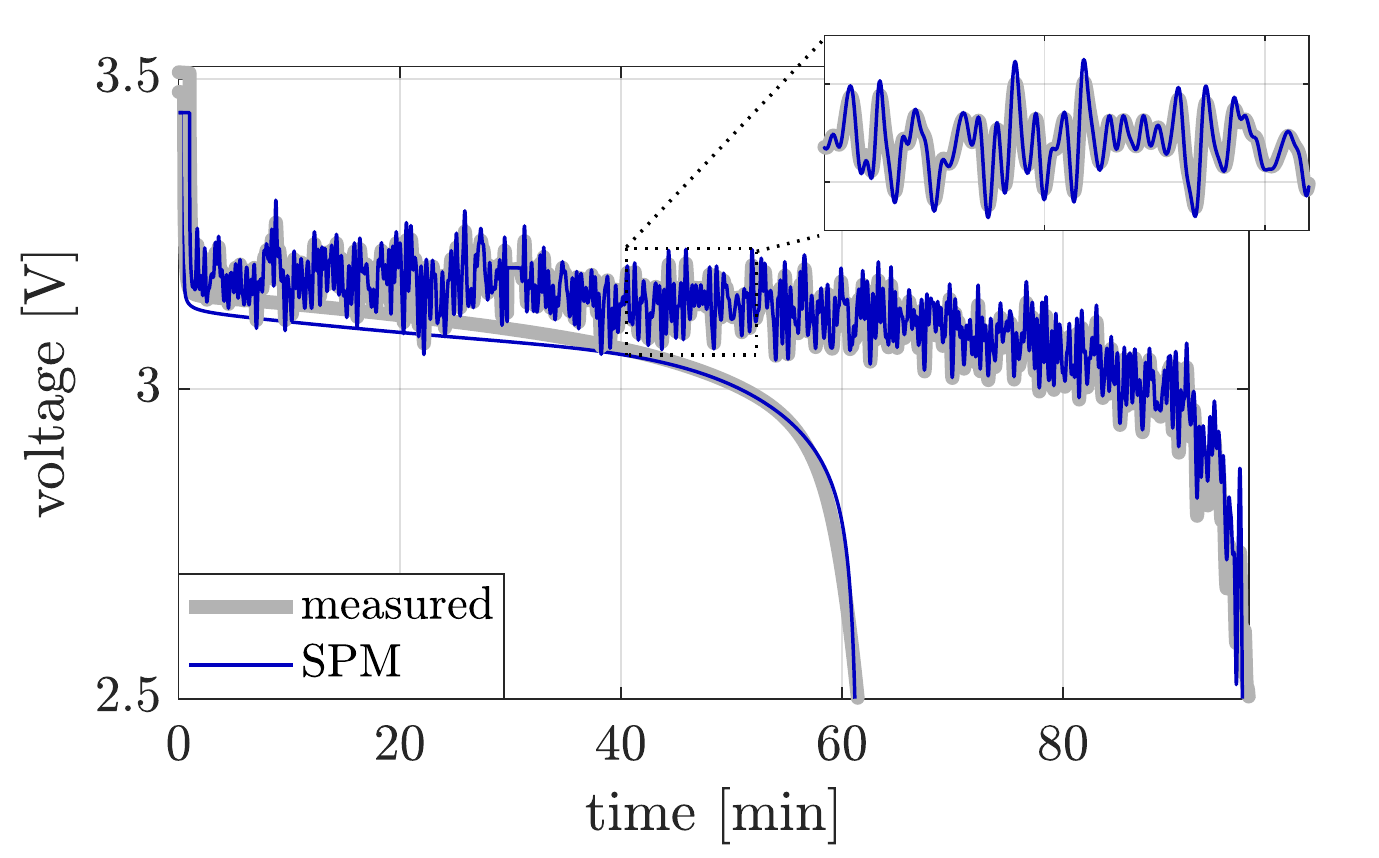}
    \caption{Measured and predicted voltage of the two identification datasets: the 1C discharge (ending around 60 minutes) and the noisy discharge.}
    \label{fitt}
\vspace{-0.6cm}
\end{figure}

\begin{table}[h!]
\normalsize
\caption{Optimal Kinetics Parameters.}
\renewcommand{\arraystretch}{1.8}
\centering
\begin{tabular}{cccc}
\hline
parameter  & \textbf{$x$*} & unit  \\ \hline \vspace{-2mm}
$V_n$    & 5.2$\cdot10^{-6}$  & $\mathrm{m}^3$  \\  \vspace{-2mm}
$V_p$   & 6.3$\cdot10^{-6}$   & $\mathrm{m}^3$ \\  \vspace{-2mm}
$r_{s,n}$   & 0.8$\cdot10^{-6}$  & $\mathrm{m}$   \\  \vspace{-2mm}
$r_{s,p}$   & 12$\cdot10^{-6}$  & $\mathrm{m}$  \\ \vspace{-2mm}
$D_{s,n}$    & 1.3$\cdot10^{-16}$  &    $\mathrm{m}^2\mathrm{s}^{-1}$ \\  \vspace{-2mm}
$D_{s,p}$    & 2.7$\cdot10^{-14}$ &    $\mathrm{m}^2\mathrm{s}^{-1}$ \\  \vspace{-2mm}
$k_n$    & 3.9$\cdot10^{-10}$  &    $\mathrm{m}^{2.5}\mathrm{mol}^{-0.5}\mathrm{s}^{-1}$  \\  \vspace{-2mm}
$k_p$  & 1.1$\cdot10^{-11}$ &    $\mathrm{m}^{2.5}\mathrm{mol}^{-0.5}\mathrm{s}^{-1}$  \\ 
$R_f$    & 38.5$\cdot10^{-3}$ &    $\Omega$ \\
\hline
\end{tabular}
\label{optKIN}
\end{table} 
\noindent The model obtained has RMSE values, for the identification datasets, of, respectively, 24 mV and 18 mV (Fig. \ref{fitt})
\newpage
\section{EQUIVALENT CIRCUIT MODEL IDENTIFICATION}
\label{s3}
\vspace{-0.1cm}
The most common Equivalent Circuit Models are composed of a SoC-driven voltage generator, modeling the OCV, a pure resistive term, modeling the instantaneous voltage drops, and a finite number of RC nets, used to model the dynamical transient responses [\cite{ecm1}]. Here, we considered a second order RC network is selected since it is typically  considered a good compromise between complexity and accuracy. Moreover, in order to enhance the ECM accuracy the model parameters are made SoC-dependant.\\
The model is described in the state-space form by:
\begin{equation}
       \begin{cases}
    \dot{V}_{c,1}(t) = -\dfrac{1}{R_1C_1}V_{c,1}(t) + \dfrac{1}{C_1}I(t) \vspace{1mm}\\ 
        \dot{V}_{c,2}(t) = -\dfrac{1}{R_2C_2}V_{c,2}(t) + \dfrac{1}{C_2}I(t) \vspace{1mm}\\
        V(t) = \mathrm{OCV}(\mathrm{SoC}) - R_0I(t) - V_{c,1}(t) -V_{c,2}(t) 
    \end{cases} 
    \label{ecmdyn}
\end{equation}

\par The current $I(t)$ is the model input, the voltage $V(t)$ is the output and the SoC is estimated via \textit{Coulomb Counting} algorithm (eq. \eqref{soc}).
The identification protocol consists of two phases: a static characterization and a dynamic identification, similarly to the SPM procedure.
\par
\begin{figure}[h!]
\vspace{-0.2cm}
\hspace{-0.1cm}
\vspace{-0.3cm}
 \includegraphics[scale=0.525]{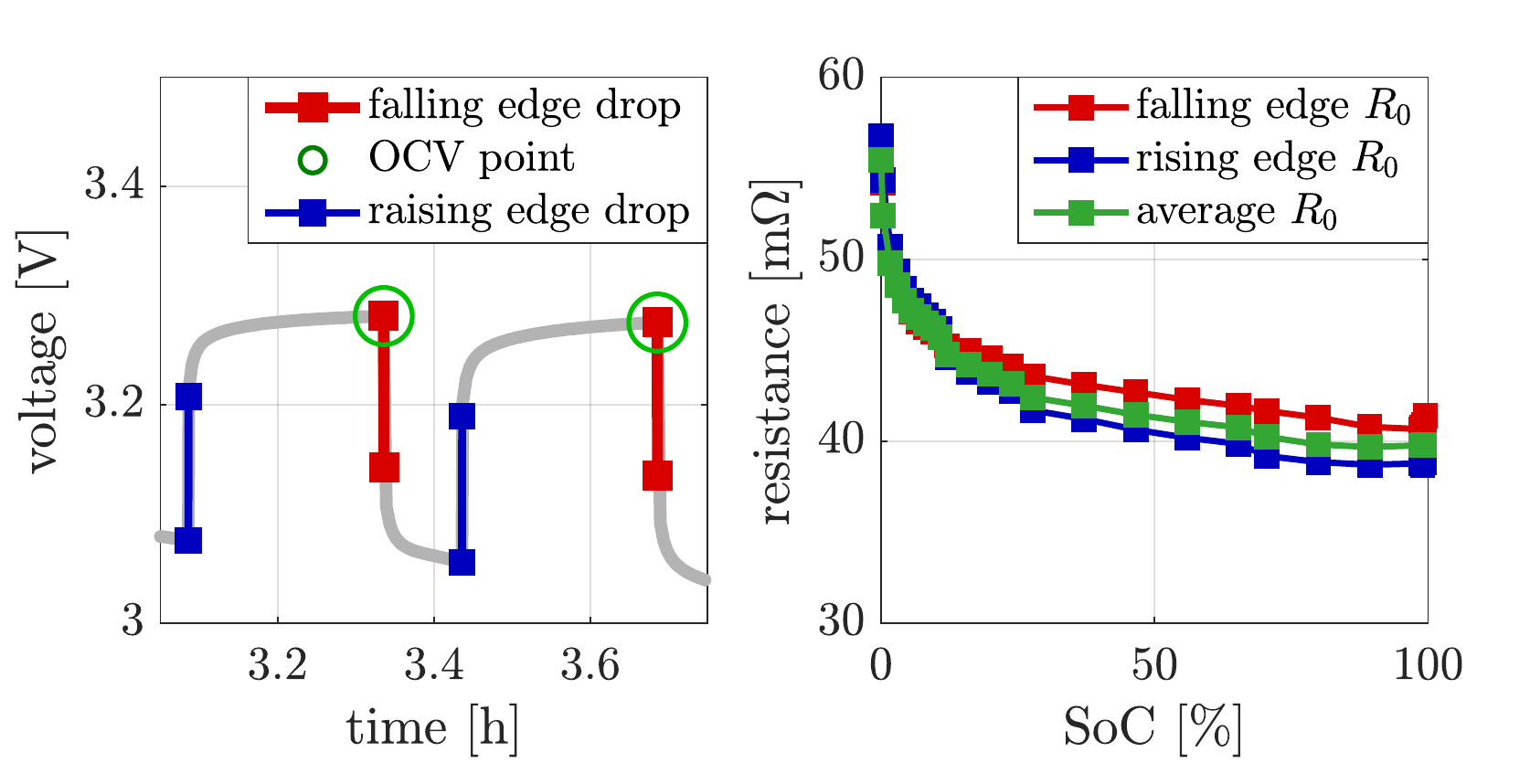}
 \vspace{-2mm}
    \caption{Detail of the Pulse Discharge Test (left) and exponential trend of internal resistance (right).}
    \label{drops}
    \vspace{0mm}
\end{figure}

 \textit{Static Characterization.} The first step consists in the identification of the OCV-SoC map and the resistance $R_0$ with a Pulse Discharge Test (PDT). Considering Fig. \ref{drops}, the voltage drop observed at application/removal of each pulse allows to estimate $R_0$, while the voltage measured at the end of each relaxation period can be approximately considered an open circuit measurement.

\noindent As visible in Fig. \ref{drops}, the resistance $R_0$ exhibits an exponential-like dependency on the SoC, in the range 40-60 m$\Omega$, and it is modelled as:
\begin{equation}
    R_0 = r_{0,1} + r_{0,2}\exp{\left(r_{0,3}\mathrm{SoC}\right)} 
    \label{socdep}
\end{equation}
\\with $r_{0,1} = 39.8$ m$\Omega$, $r_{0,2} = 12.2$ m$\Omega$ and $r_{0,3}= -6.5$, while the resulting OCV map has been already shown in Fig. \ref{ocvv} (in gray).
\par \textit{Dynamics Identification.} The remaining dynamic parameters are identified minimizing the simulation error of two training datasets, the PDT and the white noise discharge used in Section \ref{s2}.\\\\
The model can be enhanced expressing the RC parameters $R_1$, $C_1$, $R_2$, and $C_2$ as exponential functions of SoC, similarly to \eqref{socdep}:
\begin{equation}
\begin{aligned}
R_1 = r_1 + a_1\exp{\left(b_1\mathrm{SoC}\right)},\\
C_1 = c_1 + d_1\exp{\left(e_1\mathrm{SoC}\right)},\\
R_2 = r_2 + a_2\exp{\left(b_2\mathrm{SoC}\right)},\\
C_2 = c_1 + d_2\exp{\left(e_2\mathrm{SoC}\right)}.
\end{aligned}
    \label{socdep2}
\end{equation}
The optimization is solved with \verb+fmincon+ twice: with static RC parameters (ECM) and with SoC-dependant parameters as declared in Eq.\ref{socdep2} (ECM\textsubscript{+}).

\begin{table}[h!]
\vspace{-0.1cm}
\renewcommand{\arraystretch}{1.5}
\centering
\caption{Results and RMSE of dynamics identification of ECM and ECM\textsubscript{+}.}
\begin{tabular}{c  c  c }
 \hline
  parameter & ECM & ECM\textsubscript{+} \\
  \hline
  $R_1$ (m$\Omega$) & 20.9 &   $24.7+24.2\exp{\left(-6.68\cdot\mathrm{SoC}\right)}$ \\
   $C_1$ (F)& 788 & 2006 \\ 
  $R_2$ (m$\Omega$)& 21.1 & 25.7 \\
  $C_2$ (F) & 3.6$\cdot$10\textsuperscript{4} &  2.2$\cdot$10\textsuperscript{5} \\
  \hline
  RMSE in PDT (mV)  &  20.3 & 15.7 \\  
  RMSE in WN (mV) &  17.3 & 10.2   \\
\hline
\end{tabular}
\label{rmse_err}
\vspace{0cm}
\end{table}
Concerning the ECM\textsubscript{+}, the optimization problems returned that the only SoC depended parameter is $R_1$, while $R_2$, $C_1$ and $C_2$ has a nearly-flat characteristics with a negligible SoC-dependency.    
    

\section{MODELS VALIDATION AND\\ COMPARISON}
\label{s4}
Six dynamical discharge tests are used to validate the SPM and compare it with the ECM presented in Section \ref{s3} along with ECM\textsubscript{+} with the enhanced parameters. The validation tests are grouped and named according to their discharge current band: low-current tests (LC), up to 1C, medium-current tests (MC), up to 2C, and high-current (HC), up to 3C. Model performance are evaluated, once again,  with RMSE index (Table \ref{tab_val}).

\begin{table}[h!]
\renewcommand{\arraystretch}{1.5}
\centering
\caption{RMSE of the validation datasets.\\ In green: good performance (0-20 mV),\\ in yellow: acceptable performance (20-50 mV),\\ in red: poor performance (50-100 mV).}
\begin{tabular}{r| cc:cc:cc }
\hline
 test ID & LC1 & LC2 & MC1 & MC2 & HC1 & HC2   \\
   \hline
  max C-Rate & \multicolumn{2}{c:}{1C}   & \multicolumn{2}{c:}{2C}  & \multicolumn{2}{c}{3C}   \\ \hline  
           ECM (mV) &  \cellcolor[HTML]{e0ffe6} 12.5 &  \cellcolor[HTML]{e0ffe6} 18.5 & \cellcolor[HTML]{ffffcc} 23.4  & \cellcolor[HTML]{ffe3e3} 50.8  & \cellcolor[HTML]{ffe3e3} 81.3 \cellcolor[HTML]{ffe3e3}  &  \cellcolor[HTML]{ffe3e3} 141.4  \\  
             ECM\textsubscript{+} (mV) &  \cellcolor[HTML]{e0ffe6} 11.4 &  \cellcolor[HTML]{e0ffe6} 15.1 & \cellcolor[HTML]{e0ffe6} 19.1  & \cellcolor[HTML]{ffffcc} 40.9  & \cellcolor[HTML]{ffe3e3} 76.8 \cellcolor[HTML]{ffe3e3}  &  \cellcolor[HTML]{ffe3e3} 112.0  \\  
    SPM (mV) & \cellcolor[HTML]{ffffcc} 21.1  & \cellcolor[HTML]{e0ffe6} 12.5  & \cellcolor[HTML]{e0ffe6} 18.1  &  \cellcolor[HTML]{e0ffe6} 17.3 & \cellcolor[HTML]{ffffcc} 49.1  & \cellcolor[HTML]{ffffcc} 45.5  \\    
\hline
\end{tabular}
\vspace{0.2cm}
\label{tab_val}
\end{table}
\noindent It is clear that, while both models perform well at low/medium currents (0-1C/2C), they tend to lose accuracy at high current rates (0-3C).
\noindent These data demonstrate a general superiority of SPM, particularly noticeable at high current rates, where both ECM and ECM\textsubscript{+} fail to achieve acceptable performance (RMSE $\geq$ 50 mV). 

\begin{figure}[h!]
\vspace{-0.3cm}
\hspace{-0.3cm}
\vspace{-0.2cm}
 \includegraphics[scale=0.65]{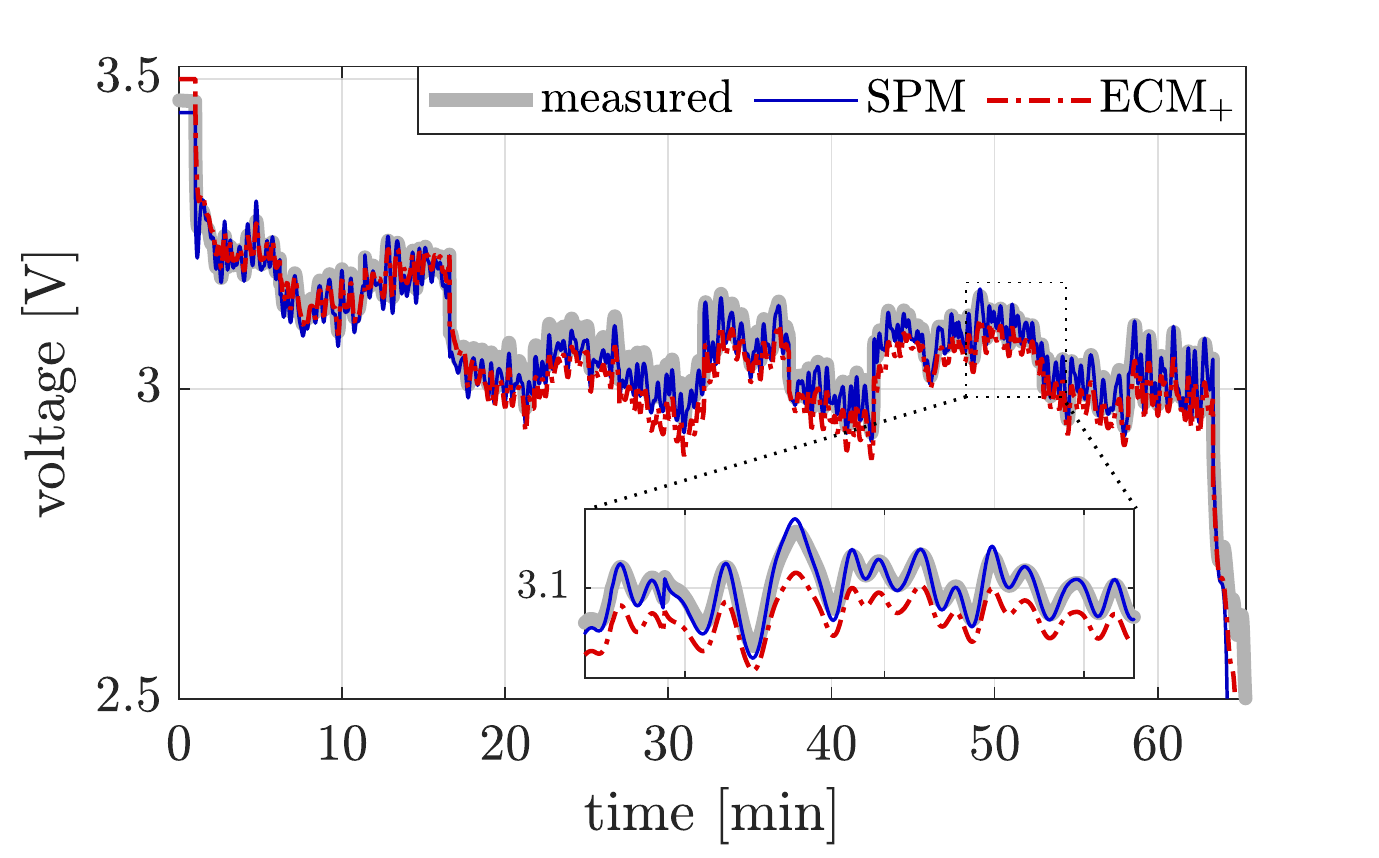}
     \caption{Measured and predicted terminal voltages of MC2.}
    \label{val1}
    \vspace{-0.1cm}
\end{figure}

For example, in Fig. \ref{val1}, the models are compared during the medium-current random discharge MC2, showing a significant drift of the ECM\textsubscript{+} prediction.

\begin{figure}[h!]
\vspace{-0.2cm}
\hspace{-0.3cm}
\vspace{-0.2cm}
 \includegraphics[scale=0.46]{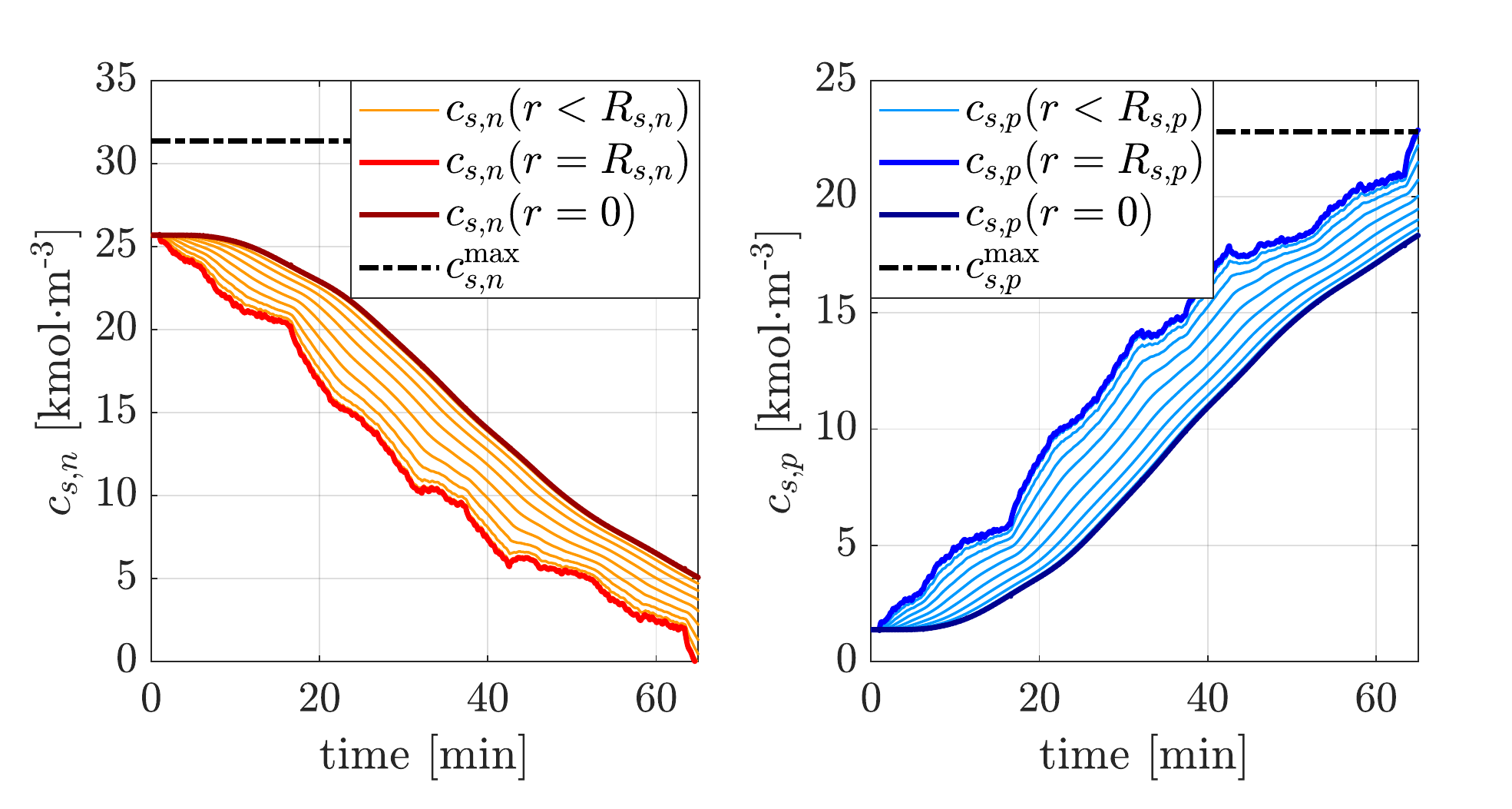}
    \caption{Anode (left) and cathode (right) solid-phase concentrations simulated with SPM during MC2}
    \label{cs}
    \vspace{-0.5cm}
\end{figure}

\noindent Besides the better model accuracy, SPM can provide helpful insights of the cell's internal states, such as the potential and the solid-phase lithium concentration of the single electrodes, the overpotentials and so on. For instance, Fig. \ref{cs} displays the solid phase concentrations of anode and cathode, along different radial coordinates, during the discharge MC2.
\vspace{2mm}
\section{CONCLUSIONS}
\label{s5}
The high number of model parameters and the non-linearity of the equilibrium potential curves of LiFePO\textsubscript{4} make the non-invasive characterization of Single Particle Model challenging when compared with traditional ECMs. We have shown that, in addition to the availability of physically-representative cell states, SPM provides more accurate voltage predictions than an enhanced ECM, especially when increasing the cell discharge rate. 

{\fontsize{7.6pt}{8.5pt}\selectfont 
\bibliography{ifacconf}             
                    }                                 






\end{document}